\documentclass[conference]{IEEEtran}
\IEEEoverridecommandlockouts
\IEEEpubid{\makebox[\columnwidth]{978-1-7281-0858-2/19/\$31.00~\copyright~2019 IEEE \hfill} \hspace{\columnsep}\makebox[\columnwidth]{ }}

\usepackage{cite}
\usepackage{amsmath,amssymb,amsfonts}
\usepackage{algorithmic}
\usepackage{graphicx}
\usepackage{caption}
\usepackage[bookmarks=false]{hyperref}
\usepackage{textcomp}
\usepackage{booktabs}
\usepackage{soul}

\usepackage[table]{xcolor}    

\definecolor{light-gray}{gray}{0.95}

\def\BibTeX{{\rm B\kern-.05em{\sc i\kern-.025em b}\kern-.08em
    T\kern-.1667em\lower.7ex\hbox{E}\kern-.125emX}}

\begin{document}

\title{Towards a Flexible System Architecture for Automated Knowledge Base Construction Frameworks}

\author{\IEEEauthorblockN{Osman Din}
\IEEEauthorblockA{\textit{Massachusetts Institute of Technology} \\
Cambridge, MA \\
https://orcid.org/0000-0001-6507-2468
}
}

%
%

\maketitle
\IEEEpubidadjcol

\begin{abstract}

Although knowledge bases play an important role in many domains (including in archives, where they are sometimes used for entity extraction and semantic annotation tasks), it is challenging to build knowledge bases by hand. This is owing to a number of factors: Knowledge bases must be accurate, up-to-date, comprehensive, and as flexible and as efficient as possible. These requirements mean a large undertaking, in the form of extensive work by subject matter experts (such as scientists, programmers, archivists, and other information professionals). Even when successfully engineered, manually built knowledge bases are typically one-off, use-case-specific, non-standardized, hard-to-maintain solutions.

Recent advances in the field of automated knowledge base construction (AKBC) offer a promising alternative. A knowledge base construction framework takes as input source documents (such as journal articles containing text, figures, and tables) and produces as output a database of the extracted information. 

An important motivation behind these frameworks is to relieve domain experts from having to worry about the complexity of building knowledge bases. Unfortunately, such frameworks fall short when it comes to scalability (ingesting and extracting information at scale), extensibility (ability to add or modify functionality), and usability (ability to easily specify information extraction rules). This is partly because these frameworks are often constructed with relatively limited consideration for architectural design, compared to the attention given to algorithmic performance and low-level optimizations. 

As knowledge bases will be increasingly relevant to many domains, we present a scalable, flexible, and extensible architecture for knowledge base construction frameworks. As a work in progress, we extend a specific framework to address some of its design limitations. The contributions presented in this short paper can shed a light on the suitability of using AKBC frameworks for computational use cases in this domain and provide future directions for building improved AKBC frameworks.

\end{abstract}


\section{Introduction}

Knowledge bases have been successfully used for various applications, including entity extraction, semantic annotation, entity disambiguation, entity resolution, question answering, and the preservation of the social Web. As an example, many sophisticated widely known applications such as IBM Watson rely on knowledge bases. Knowledge bases can also benefit, and benefit from, big data technologies: Scalable distributed big data algorithms can be used to harvest knowledge from the Web and large information repositories; the generated knowledge bases can then be used for the interpretation and processing of the ingested Web-scale big data \cite{Suchanek:2014:KBA:2733004.2733069}.

As can be expected, building knowledge bases by hand requires extensive resources. Some of the well known successful knowledge bases have taken many years to build. For example, the creation of the PaleoBioDB knowledge base took nine person-years and a group of 380 scientists \cite{zhang2015deepdive}. In addition, such knowledge bases cannot be easily reused:  one needs to reinvent the wheel and create a new knowledge base from scratch for different use cases. This presents a serious limitation for organizations with limited resources. 

A current area of investigation is automating the construction of knowledge bases, leveraging advances in big data, natural language processing, and machine learning technologies \cite{Suchanek:2014:KBA:2733004.2733069}. Automated knowledge base construction (AKBC) engines take as input a set of documents (journal articles, technical reports, etc.) and produce a database consisting of the extracted facts. For example, if there is a repository of web pages that describe various companies, given a list of named entities (such as real world names) and a relationship of interest (such as employees of the same company), an AKBC engine can output the named entities that satisfy this relationship (i.e., a list of employees by company).


There are some factors that make AKBC frameworks less than ideal, however. Among the limitations is scalability. The frameworks rely on vertical scaling (which is typically insufficient for large workloads), or, worse, leave such details to implementors. Another gap is usability. In some frameworks, users are required, for instance, to write complex rules and feature extractors in a scripting language (which requires development resources). Finally, the frameworks offer insufficient support for extensibility. As an example, there is no simple or clean method to add domain-specific features to a knowledge base framework. 

In our view, these limitations result from inadequate attention to the overarching system design or architecture, possibly because AKBC is still an active research area. Building a knowledge base construction framework is challenging, after all, as a number of components of a knowledge base (information extraction, data set creation, etc.) require attention. So far, the focus of most systems has been more on designing the underlying extraction and machine learning algorithms and making them performant, and less on system design. Indeed, we have observed that a framework is usually extracted from the specific, pressing needs of particular applications of interest, and thus such a framework may not generalize to other applications with different requirements. 

In this paper, therefore, we propose a generic framework that addresses these limitations. Our end goal is to make these frameworks easier to adopt and implement. 

The contributions of this paper are as follows:

\begin{enumerate}

\item A system architecture for a knowledge base construction framework, based on a set of general functional and non-functional requirements.
\item Extensions to a knowledge base construction framework, with the goal to align it with the proposed architecture.

\end{enumerate}

The organization of the paper is as follows. We provide background information on the use of knowledge bases in archives (and closely related communities) in section II, along with an overview of the field of automated knowledge base construction. Section III defines the requirements for AKBC frameworks. Section IV provides an overview of latest AKBC solutions and some of their limitations. Section V outlines the architecture of the proposed system. The sections that follow discuss related and future work. 

\section{Background}

\subsection{Knowledge Bases in Relevant Domains}

Knowledge bases have been used in a variety of contexts in archives and related communities. A number of examples are listed below. These projects go beyond simple information extraction techniques used in projects such as \cite{NERUsage:2018, 10.3389/fdigh.2018.00002}.

\begin{itemize}

\item The US National Library of Medicine has built a knowledge base of biomedical conference proceedings literature. The knowledge base stores information such as publications and authors. Entities extracted from a relational database are converted into a OWL-RDF based knowledge base for semantic querying. 

\item The UK National Archives (TNA) has a search system, TNA-Search, comprising of a knowledge base and a text mining tool \cite{MAYNARD12.122}. The knowledge base, built using the OWLIM semantic graph \cite{Kiryakov:2005:OPS:2113364.2113388},  contains various sources (such as resources from data.gov.uk, TNA projects, and geographical databases). Source data, comprising of government web archives, is then semantically annotated and indexed against this knowledge base, allowing semantic queries on the data. 

\item Similarly, the National Archives of France has reported creating a knowledge base comprising of various entities (government ministries and agencies, public notaries, persons, families, and private corporate bodies) \cite{french}.

\item Annotated text from The Royal Commission on the Ancient and Historical Monuments of Scotland (RCAHMS) has been used by a European project to create a querable graph of nested name entities \cite{4338398}.

\item Finally, the ARCOMEM (European Archives, Museums, And Libraries in the Age of Social Web) project has a knowledge base of archived web content \cite{dietze2012entity}.  The goal of the project is to “preserve archived Web content over time and allow its navigation and analysis based on well-formed structured RDF data about entities.” 

\end{itemize}

To the best of our knowledge, existing literature does not provide evidence on the use of automated knowledge base frameworks for archives and related domains. 


\subsection{Automated Knowledge Base Construction} 

Knowledge base construction is the process of populating a database with information from text, tables, images, video, and even incomplete knowledge bases. 

Examples of AKBC frameworks include Fonduer \cite{wu2017fonduer}, DeepDive \cite{zhang2015deepdive}, MinIE \cite{gashteovski2017minie}, Alexandria \cite{anonymous2019alexandria:},  NELL \cite{NELL-aaai15}, KnolwedgeVault \cite{45634}, etc. Examples of automatically populated knowledge bases, comprising of real world entities such as people and places, include, YAGO, Freebase, DBPedia, YAGO2, and Google Knowledge Graph \cite{45634}.


To understand how these knowledge base frameworks work, it is important to note that different knowledge bases have processing pipelines that comprise different phases.

The first phase is candidate generation and feature extraction. In this phase, pre-processing NLP tools (entity tagging, for instance) are applied, and candidate features are extracted from the text, based on user defined rules. Some frameworks that rely on a generative model of text (such as Alexandria) may include a pre-processing stage but do not have a feature extraction phase.

Next comes the supervision and classification phase, and this is where some form of training data is used. The training data can be manually labelled or it can be created through techniques such as distant supervision (whereby an existing database, such as Freebase, is used) and weak supervision (whereby training data is programmatically generated). Unsupervised systems such as Alexandria do not require training data. 

The supervision phase is followed by a learning and inference phase, where models such as LSTM (a type of a recurrent neural network that can capture long-term dependencies in a text) are used. Some systems have an analogous statistical inference phase, whereby a schema is derived using inference rules or a probabilistic model (such as a Markov Logic Network). 

Finally, some knowledge base frameworks include an error analysis step, whereby information from previous phases can be used to correct extraction mistakes or inaccurate features. Unsupervised systems may omit this phase altogether. 

\section{Architecture Requirements}

Besides underlying requirements for precision, high coverage, and being up-to-date, the main requirements of the system are as follows.

\subsection{Functional Requirements}

\begin{enumerate}
\item \textbf{Support for multiple types and formats of data}. AKBC frameworks must offer the capability of processing a diversity of data and data formats.
\item \textbf{Support for storage and search}. The knowledge base framework must store extracted facts in a format that is indexable and queryable.

\item \textbf{Support for flexible feature selection.} To allow for variation and noise in input text, extraction rules should be flexible, and not rigid expressions or regex-like patterns. 

\item \textbf{Support for adding domain features}. As there is variation between corpora from different domains, it must be possible to add domain-specific features to a knowledge base construction framework to increase the accuracy and completeness of a knowledge base.

\item \textbf{Support for human feedback}.  For systems that require any user input, the knowledge base framework should support error analysis to fix (or flag) incorrect or overly-specific features. 

\end{enumerate}

\subsection{Non-functional Requirements}

\begin{enumerate}

\item \textbf{Performance}. The system should be performant when training a model or applying inferences.

\item \textbf{Scaling}. The system must be able to scale in order to process a large corpus of potentially billions of documents, containing, in turn, billions of figures and tables. This is increasingly relevant as larger and larger data sets become available, and especially as new forms of archives, social and web, emerge.

\item \textbf{Usability}. The system must not require end users to learn technical details of underlying algorithms. The system should also not require writing complex extraction functions (in the form of programs or scripts).

\item \textbf{Support for transparency and fairness}. The system should provide the capability to choose between different features (and even models), as this can allow end users to decide if any features or models do not meet desired properties (such as fairness).

\end{enumerate}

We recognize that an organization that implements a knowledge base system is likely to have application-level requirements for maintainability, security, and system management. Such practical, implementation-level requirements are out of the scope for our discussion.

\section{State-of-the-art Overview and Limitations}
In this section, we briefly introduce a number of modern knowledge base construction engines and discuss their limitations. For brevity and for adherence to some of the requirements (in particular, flexible feature selection), we restrict our attention to machine learning based systems (as opposed to rule-based systems). We also do not consider open information extraction systems (such as MinIE), as they are more prone to errors (such as duplicate facts due to slight changes in wordings).

\subsection{Overview}
\textbf{Fonduer} \cite{wu2017fonduer} is a knowledge base framework concerned with richly formatted data (prevalent in web pages, business reports, product specifications, etc.), where relations and attributes are expressed via combinations of textual, structural, tabular and visual information. The key insight behind Fonduer’s extraction approach is that the organization and layout of a document determines the semantics of the data.  To represent features of relation candidates, Fonduer uses a bidirectional Long Short-term Memory (LSTM)  with attention. Relying on LSTM, along with weak supervision, obviates the need to create large sets of training data by hand, an important consideration since it is difficult to build proper training data at a large scale.

\textbf{DeepDive} \cite{zhang2015deepdive}  uses manually created feature extractors to extract candidate facts from text. In addition to manually labelled training examples, DeepDive supports distant supervision. This allows a user to define a mapping between a preexisting, yet incomplete, knowledge base (possibly manually created) and a corpus of text. DeepDive uses Markov Logic Networks (MLN), a probabilistic model.


\textbf{Alexandria} \cite{anonymous2019alexandria:} also makes use of a probabilistic machine learning model. Alexandria creates a probabilistic program, which it inverts to retrieve the facts, schemas, and entities from a text. Alexandria does not require any supervision (only a single seed example is required). 


\rowcolors{1}{white}{light-gray}
\begin{table*}[h]
\begin{center}
\setlength\extrarowheight{1.8pt}
\begin{tabular}{p{3cm} p{3cm}  p{3cm}  p{3cm}}
\textbf{Requirement} & \textbf{Fonduer} & \textbf{DeepDive} & \textbf{Alexandria} \\[15pt]

RQ1. Support for multiple types and formats of data & Multiple formats are supported. & Multiple formats are supported (including limited supported for image extraction). & Textual Web data is supported. \\ 


RQ2. Support for storage and search & Relational database & Relational database & N/A \\[15pt]


RQ3. Support for flexible feature selection & Weak supervision & Distant supervision & Automated extraction \\[15pt]

RQ4. Support for adding domain features & No specific support. & No special support. & N/A \\[15pt]


RQ5. Support for human feedback & Reliance on declarative programming. & Error feedback phase facilitated by SQL queries & N/A \\[15pt]

RQ6. Performance optimizations (support for minimizing training time) & Low-level optimizations (such as caching). & Low-level techniques (such as buffer replacement) & Low-level optimizations (such as caching). \\[15pt]


RQ7. Scaling (support for large data sets) & No specific support. & Low-level optimizations and vertical scaling. & Support for Microsoft COSMOS distributed computing and Azure Batch framework \\[15pt]


RQ8. Usability & No specific support. & No specific support. & N/A \\[15pt]


RQ9. Support for transparency and fairness & No support. & No support. & No support. \\[15pt]


\end{tabular}
\end{center}
\caption{Relation between different knowledge base construction frameworks and requirements presented in Section III.}
\label{tab:my-table}
\end{table*}

\subsection{Missing Features}

Table I illustrates how the selected AKBC engines adhere to the requirements postulated in Section III. We observe that the selected engines fulfill some of the requirements, but there are a number of instances where the design of these systems does not adequately address the requirements:

\begin{enumerate}

\item First, the selected frameworks do not provide the functionality to easily add domain features. This limitation may therefore mean that end users have to hard code such features into the framework itself.

\item Second, the frameworks do not allow their pipelines to be extended easily. This may result in burdening end users with updating the framework source code directly to add certain phases (to process images, for example).

\item Third, the frameworks do not offer a user interface. This may require users to add extraction rules or exclusion filters through a special program or script they have to write (requiring some knowledge of a scripting language).

\item Finally, the frameworks offer limited support for the scale-out architectural pattern. This may prevent the system for scaling, and its performance may degrade as more content is ingested. 

\end{enumerate}


As these features are missing or inadequately addressed in the selected AKBC frameworks, the frameworks fall short in their intent (often stated explicitly, as in \cite{zhang2015deepdive}) of lessening the burden of domain experts. The end user is still required to do considerable heavy lifting, by learning the internals of a framework and extending it. The goal of this work, therefore, is to fill this gap by postulating an architecture that takes these considerations into account and arguing that these changes must be accounted for by framework designers.

\section{System Architecture}

In this section we outline a proposal for an architecture for knowledge base frameworks. First, we introduce a number of design principles that guide the architecture. Next, we describe an architecture that is consistent with these principles and the set of functional and non-functional requirements from Section III. Finally, we describe a preliminary realization of this architecture, in the form of extensions to a specific knowledge base framework.

\begin{figure*}[hbt!]
\label{appdiagram}
\centering
\includegraphics[scale=0.65]{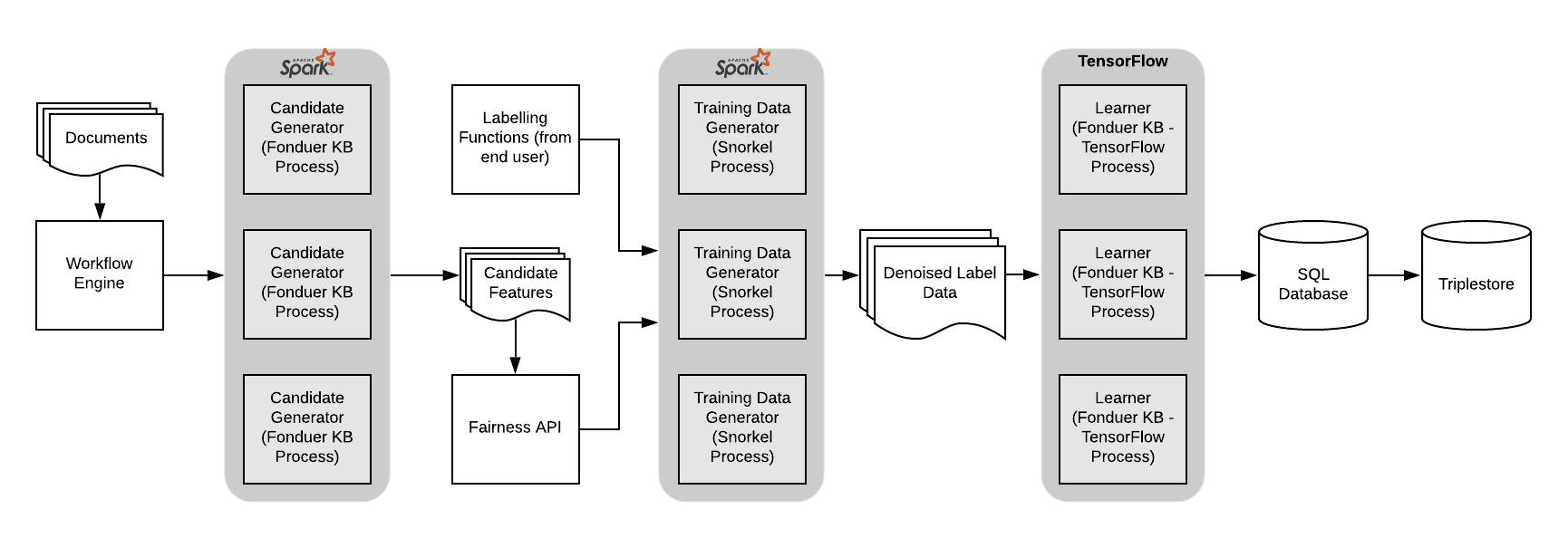}
\caption{Overview of the system architecture. The API-based design allows for horizontal scalability, usability, and extensibility.} 
\end{figure*}

\subsection{Design Principles}
Before discussing the architecture, we establish a number of key design principles.

\begin{enumerate}
\item \textit{The framework's design should be based on APIs}. Exposing the underlying functionality through APIs can make it easier to scale and customize the framework in accordance with difference use cases.
\item \textit{Middleware services should be used}. Leveraging middleware services instead of point-to-point connections between components of the system can make it easier to rapidly implement new use cases and functionality.
\item \textit{The design should not be reliant on proprietary components}. Among other factors, depending on proprietary vendor solutions can result in unsustainable solutions. 
\item \textit{Transparency and fairness aspects should be weighed}. Filtering out discriminative information (often a negative consequence of machine learning systems) at the very source of data generation can prevent biases in upstream applications.
\end{enumerate}

\subsection{Architecture}

The system consists of several components, as illustrated in  Figure 1. The major components of the system are listed below.

\begin {enumerate}
\item \textbf{Knowledge base framework}. The core of the system is the knowledge base construction engine.
\item \textbf{Distributed middleware}. Different components of the framework are scaled out using TensorFlow (a machine learning library) and Apache Spark (a cluster computing framework). Leveraging these solutions enables distributed model training and distributed supervision.
\item \textbf{Persistence middleware}. A middleware component allows the replication of extracted relations in the database to a triple store (after transformation into RDF). A relational database enables ACID-based transactions, while a triple store facilitates upstream RDF based applications. 
\item \textbf{Graphical user interface}. A dedicated user interface allows end users to provide extraction rules and filters in a user friendly way. The interface also provides a summary of feature candidates flagged for review. 
\item \textbf{Workflow engine}. An external workflow engine (such as Airflow) is responsible for programmatically ingesting documents from different source repositories.
\end{enumerate}

\subsection{Architectural Extensions to Fonduer}

To demonstrate our position, we describe the application of these architectural building blocks to a knowledge base framework. The system, named System Architecture for the Generation and Extension of Knowledge Bases (SageKB), is a work in progress. All artifacts are available on the project’s website \cite{sageKB}. 

The following are the main components of the system:

\begin{enumerate}

\item \textbf{Fonduer knowledge base engine}. We selected Fonduer over related frameworks, as Fonduer can extract information from richly formatted data, not just textual data. Compared to Fonduer, a limitation with DeepDive is that it is no longer being actively developed, and the project itself considers Snorkel-based and Fonduer-like approaches to be its successors \cite{deepdiveweb}. A limitation of Alexandria is that it is a work in progress and details about the system are currently unknown. A limitation of selecting Fonduer is that it lacks the capability of extracting data from figures, an important source of information in scholarly documents and other publications (such as reports and presentations). We think that an API-based approach will allow us to add other extraction algorithms as needed, as part of the pipeline.  The approach may even make it possible to use an ensemble of different algorithms under a single framework. Finally, the ideas described here can be applied to other knowledge base frameworks, and they are not restricted to a specific framework or a particular class of frameworks. 
\item \textbf{APIs}. Since Fonduer lacks a web service API, we added an API to the framework.  This is a first step towards scaling Fonduer. An API also makes it easier to expose critical functionality via a graphical user interface, instead of requiring the end user to make changes to the framework source code itself.  
\item \textbf{Distributed training through TensorFlow}. The TensorFlow library allows distributed training, making it possible to ingest a large, Web scale repository. Models can be trained on many machines (using horizontal scaling), as opposed to a single powerful machine (vertical scaling).  
\item \textbf{Distributed weak supervision through Apache Spark and Snorkel}. Apache Spark allows Snorkel \cite{snorkel} processes, used for weak supervision, to be distributed to many nodes, thus reducing the time for learning.  
\item \textbf{Integration with a fairness API}. A separate API helps determine if any of the generated candidate features are discriminative.  An example is a scenario where a table in a source document lists neighborhoods in a city and associated crime rates, and a separate table lists neighborhoods and ethnic backgrounds of its residents. A discriminative relation that may end up in the knowledge base could be residents of a particular ethnic background more likely to commit crimes than residents of other ethnic backgrounds. To prevent this, potentially discriminatory features (such as ethnic background) can be monitored and flagged (and if necessary, rejected) by an end user. This novel extension ensures that upstream machine learning applications have less imbalanced source data. 
\end{enumerate}

\section{Related Work}

There are several machine learning based frameworks and algorithms that extract content from figures (such as graphical plots) that are prevalent in scholarly works  \cite{10.1007/978-3-319-46478-7_41, 10.1007/978-3-319-71249-9_9}. These systems suffer from similar architectural limitations as the AKBC frameworks we have discussed and do not adequately address system design issues.

In their work, Choudhury and Giles (2015) propose a modular architecture for analyzing figures that are present in scholarly documents \cite{RayChoudhury:2015:AIE:2740908.2741712}. The design consists of separate modules for extracting figures and for using NLP techniques to understand the semantics of these figures. Their design approach is certainly an improvement over monolithic architectures, where functionality is built as one giant “ball-of-mud,” as is the case with most current knowledge base engines. Our approach differs from \cite{RayChoudhury:2015:AIE:2740908.2741712} in that besides modularity, it addresses the concerns of scalability and usability.

 AKBC frameworks such as DeepDive partly rely on performance characteristics of relational databases or use low-level performance techniques \cite{zhang2015deepdive}. For feature extraction, Hadoop and Candor frameworks have also been used \cite{niu2012deepdive}. Similarly, Alexandria relies on low-level techniques, and can make use of a big data framework \cite{anonymous2019alexandria:}, which in Alexandria's case is proprietary. Our work differs from these projects in that we incorporate scalability aspects into the framework itself and in that we make explicit use of horizontal scalability, realized through the use of open source frameworks such as Apache Spark. In addition, the discussed selected works do not directly address usability and extensibility aspects.


\section{Conclusions and Future Work}

In this paper, we have presented a flexible architecture for frameworks that aim to provide end-to-end functionality for automating the development of knowledge bases. We identified some of the challenges associated with AKBC frameworks and postulated a set of important requirements. Based on these requirements, we discussed the features missing in these systems, and presented an architecture that is more scalable, usable, and extensible than current approaches. We also presented an initial implementation of our ideas, in the form of extensions to a knowledge base engine. 

In terms of future work, a logical next step will be creating a user interface that leverages the system API, likely resulting in a less steep learning curve for end users. Another further direction is investigating the set of domain features (from an archives use case), with the goal to increase the precision and coverage of the knowledge base. We plan to share our implementation experience in the form of a case study. 

Finally, an implicit contribution of this paper is raising awareness of the potential of AKBC frameworks and pointing out a number of current limitations. The presented ideas should therefore prove useful to builders and end users of AKBC engines alike. As AKBC is an active area of research, we hope to share our experiences and feedback with the AKBC community \cite{akbc-conference}, highlighting areas for future investigation and improvement, from the perspective of computational use cases from this domain.


\bibliography{bib} 
\bibliographystyle{IEEEtran}


\end{document}